\documentclass{article}
\usepackage{emulateapj}
\usepackage{rotating}
\usepackage{lscape}
\usepackage{flushrt}
\lefthead{Henriksen {\em et al.}}
\righthead{Re-constructing the Merger History of the A3266 Galaxy Cluster}

\slugcomment{accepted by {\em The Astrophysical Journal}, August 1999}
\begin{document}

\title{Re-constructing the Merger History of the A3266 Galaxy Cluster}

\author{Mark Henriksen}
\affil{Physics Department, University of Maryland, 1000 Hilltop Circle, 
Baltimore MD, 21250}
\centerline{henrikse@umbc.edu}
\vskip 0.1in
\author{R. Hank Donnelly}
\affil{Harvard-Smithsonian Astrophysical Observatory, 60 Garden Street, Cambridge MA, 02116}
\centerline{rdonnelly@cfa.harvard.edu}
\vskip 0.1in
\author{David S. Davis}
\affil{MIT,Center for Space Research, Bldg. 37-662B, Cambridge MA, 02116}
\centerline{dsd@pikaia.mit.edu}
\begin{abstract}

A temperature map of the A3266 galaxy cluster has been derived from
the {\it ASCA} GIS observations. It shows an asymmetric pattern of
heating indicative of an ongoing merger between a group sized
sub-cluster and the main cluster. The galaxy distribution shows two
peaks connected in a barlike structure running NE to SW through the
central region of the main cluster, defining the merger axis. The
temperature of the intergalactic medium generally decreases from SW to
NE along the merger axis with a peak of 13.2 +3.4/-2.0 keV in a region
which is perpendicular to the merger axis and extends through the main
cluster density peak.  The central bar has a velocity dispersion of
$\sim$1300 km s$^{-1}$, compared to $\sim$1000 km s$^{-1}$ for the
surrounding cluster.  The central bar also has two distinct density
peaks in the distribution of galaxies, yet it has a velocity
distribution which is consistent with a single Gaussian. This implies
a merger in the plane of the sky. The optical and X-ray data taken
together show that a loose group of about $\sim$30 galaxies has
penetrated the main cluster from the SW, decoupling from their
original intergroup medium and passing through a strong shock
front. Two radio galaxies, one a wide angle tail morphology (WAT) and
the other a narrow angle tail (NAT), are located to the SW of the main
cluster in the post-shock gas. Since the merger is in the plane of the
sky, a dynamical analysis cannot be applied to derive the velocity of
the merger. Alternatively, using the pre- and post-shock gas
temperature and assuming an adiabatic shock, we calculate a relative
gas velocity of $\sim$1400 km s$^{-1}$.  The alignment of the NAT and
WAT relative to the shock front combined with the high gas velocity
from the merger implies that the observed radio morphologies formed
via ram-pressure as a result of the merger.
\end{abstract}

\section
{Introduction} One of the significant contributions to emerge from
{\it ROSAT} and Einstein Observatory studies of galaxy clusters is the
discovery that many clusters contain substructure or asymmetric
morphologies. The frequency of substructure, which has been estimated
at 40\% (Forman \& Jones 1990), suggests that clusters are still
forming hierarchically through merger and accretion of subclusters and
galaxy groups.  While the prevalence of substructure has primarily
been used to constrain cosmologies (Richstone, Loeb, \& Turner 1995;
Mohr et al. 1995), detailed studies of merging clusters impact a
number of important areas including: the radio properties of galaxy
clusters, cooling flow formation and evolution, galaxy evolution, and
gravitational mass measurements.

An analysis of the galaxy kinematical data can be used in many cases
to determine the geometry and dynamics of the cluster merger. While
some clusters, such as A548 (Davis et al. 1995), may have multiple
components, for many a simple 2-body dynamical analysis is sufficient
and provides an estimate of the relative velocity of the infalling
substructure.  Typical velocities are high: $\sim$2000 km s$^{-1}$ for
A400 (Beers et al. 1992), $\sim$2000 km s$^{-1}$ in A754 (Zabludoff \&
Zaritsky 1995), $\sim$3000 km s$^{-1}$ in A2256 (Roettiger, Burns, \&
Pinkney 1995), $\sim$1600 km s$^{-1}$ in A3395 (Henriksen \& Jones
1996), and $\sim$2500 km s$^{-1}$ in A2255 (Burns et al. 1995).  Since
these values indicate that subclusters merge at Mach $\sim$ 2-3, it is
expected that the merging atmospheres should show evidence of shock
heating.

{\it ASCA} observations are in a unique position to provide a physical
description of shock heating of the ICM during a merger because the
data combine both spectral and spatial resolution.  Temperature maps
of clusters with substructure sometimes show complex structure (e.g.,
Markevitch 1996) and have proven to be a powerful diagnostic of the
dynamical state of the (ICM).  For example, the two-dimensional
temperature maps of A754 (Henriksen \& Markevitch 1996), A1367
(Donnelly et al. 1998), and the Coma cluster (Honda et al. 1996;
Donnelly et al. 1999) have generally confirmed the accuracy of
hydrodynamical simulations of cluster mergers (Evrard, Metzler, \&
Navarro 1996: Roetigger, Stone, \& Mushotzky 1998; Gomez et al. 1997).

In this paper, we present an X-ray and optical analysis of A3266 using
{\it ASCA} and {\it ROSAT} observations along with positions and
velocities of over 380 galaxies taken from Quintana, Ramirez, \& Way
(1996).  A3266 is an ideal cluster in which to study the effects of a
merger using {\it ASCA}. It has a large angular radius ($\sim$ 13
arcmin) like the A1367, Coma, and A754 clusters so that several
regions can be analyzed while systematic uncertainties associated with
modeling the ASCA point-spread-function (PSF) are minimized.  Like
A754, it is very bright with a 2-10 keV luminosity of $\sim$10$^{45}$
ergs sec$^{-1}$ so that the statistical uncertainties in the
spectro-spatial analysis are minimized.

\section{Observations}

A3266 has z = 0.0594 (Quintana, Ramirez, \& Way 1996) and located at
4:31:0.9, -61:28:42 (J2000).  A3266 was observed with {\it ASCA} for
23,632 seconds on 1995 October 7-8 and with the {\it ROSAT} PSPC for
13,547 seconds in the fall of 1993.  Good {\it ASCA} data was obtained
with both the Gas Imaging Spectrometer (GIS), which has two detectors,
and with the Solid State Imaging Spectrometer (SIS), which consists of
two camera assemblies. The GIS has higher effective area at higher
energy ($>$ 5 keV) than the SIS so using both data sets are
potentially useful in analyzing multi-temperature emission. However,
A3266 was observed with the SIS in 1-CCD mode and the cluster more
than fills the chip. To maintain consistency in our analysis, only the
GIS data was used for this study.  The data were filtered using the
REV2 criteria utilized by the {\it ASCA} Guest Observer Facility in
reprocessing data since June 1997. These criteria exclude data
obtained under the following conditions: with a radiation belt monitor
(RBM) count $>$ 100 cts/s, during earth occultation or at low
elevation angle to the Earth ($<$ 5 degrees for the GIS), when the
pointing was not stable (deviation of $>$ 0.01 degrees), during South
Atlantic Anomaly passage, and when the cutoff rigidity (COR) was $>$ 6
GeV/c.

\section{Analysis}

The spectra from the {\it ROSAT} PSPC and the {\it ASCA} GIS2
detectors are first analyzed to obtain an emission weighted
temperature for the cluster. Large regions were chosen for each data
set to be co-spatial and contain all of the cluster emission; bright
point sources were removed from both images.  The PSPC region is a
circle of radius 16.3' located at 4:31:19.02, -61:27:32.98
(J2000). The GIS region is a circle of radius 17.25' located at
4:31:41, -61:26:32.96 (J2000).  The PSPC and GIS data are fit
separately and jointly with a Raymond \& Smith spectrum. The PSPC has
an energy range of 0.1 - 2.0 keV while the GIS has a broader band, 0.7
- 11 keV. Fitting both data sets separately is useful for analyzing
the presence of multiple temperature components since a difference in
observed emission weighted temperature implies multiple temperature
components are present.  In these fits, the column density,
temperature, and abundance are free parameters. The normalization for
each temperature component in each data set are also free parameters.
The $\chi^{2}$ test is used to find the best fit values and the 90\%
confidence limit on the temperature.

The primary purpose of analyzing the {\it ASCA} spectrum is to
spatially resolve temperature components implied by multi-component
modeling of the integrated spectrum.  A moderate spatial resolution
temperature map is prepared to indicate the general temperature
structure of the cluster. This map is then used to select eight
regions in which temperature are measured with greater accuracy using
the PSF modeling technique described in Churazov et al. (1996).  The
energy range covered is 1.5-11 keV.  Data below 1.5 keV and between
2.0 - 2.5 keV is not used to minimize the uncertainty in the PSF
correction (see Markevitch 1996 for details).

For {\it ASCA} observations at high Galactic latitude, blank sky
observations obtained with long exposures are utilized for background
subtraction.  The blank sky observations were taken with different COR
values.  Since the A3266 observations and background observations are
taken at different times the range of COR values are different during
each observation. Therefore, blank sky observations at each COR value
are time weighted to make a composite background image which is then
subtracted from the cluster image.  There is a day-to-day variation in
the GIS background of 20-30\% for a specific COR value, however, the
composite background image is more accurate and the uncertainty in the
background normalization is estimated at 5\% (Markevitch 1996 and
references within).  In all of the regions modeled using the PSF, the
column density is fixed at the Galactic value, 3$\times$10$^{20}$
cm$^{-2}$ and the abundance at an average value of 0.30 Solar; only
the temperature of each region is a free parameter.

\section{Results}
\begin{deluxetable}{ccccc}
\footnotesize
\tablewidth{0pt}
\tablecaption{Spectral Modeling Results}
\tablehead{
\colhead{Data Set}& \colhead{$\chi^{2}_{\nu}$} & \colhead{N$_{H}$}(cm$^{-2}$) & \colhead{kT (keV)}&
 \colhead{Abundance} 
}
\startdata
ROSAT & 0.91 & 0.023 - 0.026 & 4.7 - 6.9 & 0.00 - 0.50 \nl
ASCA  & 1.1 & 0.0023 - 0.039 & 7.7 - 8.8 & 0.14 - 0.25  \nl
ROSAT \& ASCA (1RS)	& 1.05 & 0.0225 - 0.0241 & 7.75 - 8.46 & 0.15 - 0.25 \nl
ROSAT \& ASCA (2RS)	& 1.04 & 0.0225 - 0.0234 & 0.31 - 6.5 (7.82 - 12.94) & 0.14 - 0.24 \nl
\enddata
\label{tab:tab1}
\end{deluxetable}

\subsection{Optical}

\begin{figure}
\centerline{\includegraphics[height=6.0in]{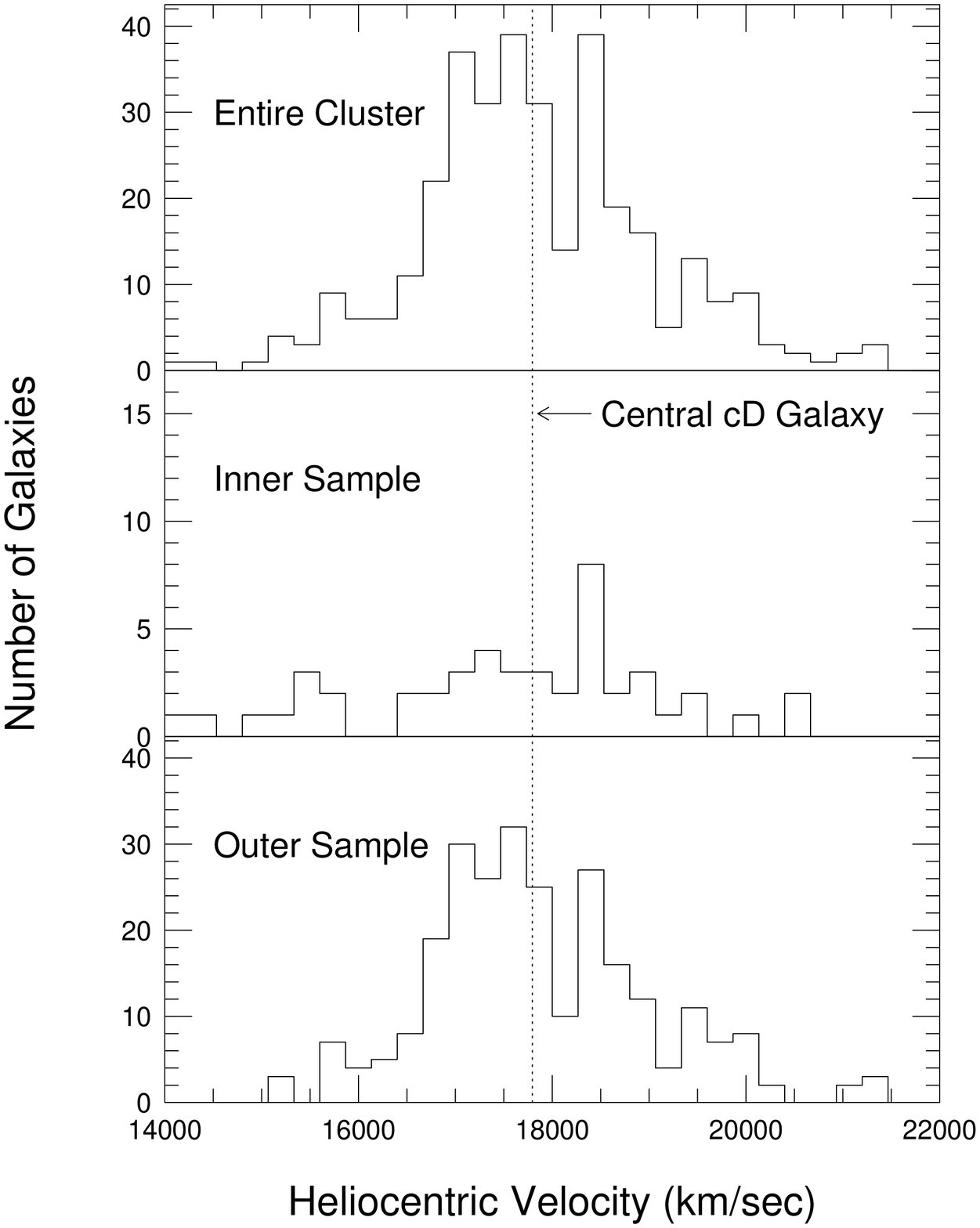}}
\caption{The top panel shows the velocity histogram after
clipping the velocities in Table 1 of QRS. There are 336 galaxies
between 14000 and 22000 km s$^{-1}$. The KMM algorithm is used to
partition the entire sample. The partitioning results in two
distributions, an inner and outer region. The inner region has a
dispersion of $\sim$1300 km sec$^{-1}$ and contains a the bar-like
structure seen in Figure~\ref{fig:fig2}. The lower panel shows the outer region
which has a velocity dispersion of $\sim$1000 km
sec$^{-1}$.}\label{fig:fig1}
\end{figure}

\begin{figure}
\centerline{\includegraphics[width=6.0in]{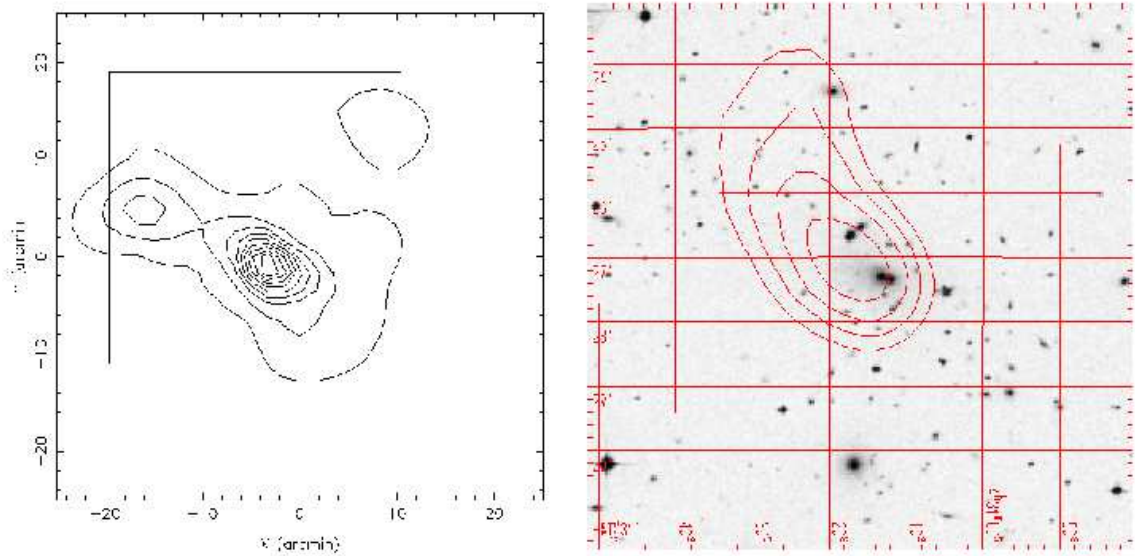}}
\caption{The left panel shows the optical image smoothed with an
adaptive kernel.  The contour levels are between 0.12 and 1.16
galaxies arcmin$^2$ with 9 contour levels separated by 0.12 galaxies
arcmin$^2$. The center of the image is J2000 4:31:01.91 -61:29:00.78.
The box in the left panel is X-ray analyzed region shown in Figure~\ref{fig:fig3}.
The right panel shows ROSAT contours overlayed on the Digitized Sky
Survey image of the central 8.5x8.5 arc min of the A3266 cluster. The
contours are derived from the ROSAT PSPC image convolved with a
1.5$\sigma$ Gaussian and range from 30 to 24 $\sigma$ in increments of
2.}\label{fig:fig2}
\end{figure}

Table 1 of Quintana, Ramirez, \& Way (1996) contains positions and
velocities for over 380 galaxies.  After clipping the velocities in
their Table 1 we retain 336 galaxies between 14000 and 22000 km
s$^{-1}$.  The top panel of Figure~\ref{fig:fig1} shows the velocity histogram for
the entire sample of 336 galaxies.  We have used the KMM algorithm to
characterize the optical morphology.  The KMM ("Kaye's" Mixture Model)
algorithm fits a user specified number of Gaussians to a dataset using
the EM method (Dempster, Laird, \& Rubin 1977) and evaluates the
improvement of that fit over a single Gaussian. It also provides a
maximum-likelihood estimate of the membership of the data points to
specific groups. For this implementation of the algorithm we input an
initial guess for the number of groups, their location (RA, DEC and
velocity) and size.  The algorithm then iteratively determines the
best position for the Gaussians and assigns each point a probability
of being a member of one of the groups. This iterative process
continues until a stable solution is found.  The KMM algorithm divides
this cluster into two groups, an inner region, which contains the bar
seen in the spatial distribution of the galaxies (see Figure~\ref{fig:fig2}), and
an outer region which is more diffuse and spans the cluster.  Shown in
the middle panel of Figure~\ref{fig:fig1} is the velocity histogram for the inner
bar-like region.  The bottom panel of Figure~\ref{fig:fig1} shows the velocity
histogram for the surrounding cluster.  The w and Anderson-Darling
(AD) statistical tests address the hypothesis that the cluster
distribution is drawn from a parent population which is a single
Gaussian (Beers et al. 1991, Yahil \& Vidal 1977).  Both tests reject
the hypothesis that the cluster velocity distribution can be drawn
from a Gaussian distribution at 90\%, and 99\% confidence
respectively.  The biweight estimate of the velocity dispersion is
1367 km s$^{-1}$ for the inner part and 1007 km s$^{-1}$ for the outer
part. This matches the velocity dispersion estimate of Quintana et
al. of 1400 km s$^{-1}$ for the inner 10 arcmin. These authors also
reported a drop in velocity dispersion from 1307 km s$^{-1}$ within
2.5h$_{50}^{-1}$ Mpc to $<$800 km s$^{-1}$ outside of 3.0h$_{50}^{-1}$
Mpc.

The spatial distribution of galaxies is shown in a contour map
prepared using an adaptive kernel (see Figure~\ref{fig:fig2}, left panel). To
create this map, an initial smoothing scale of 0.25 Mpc is chosen and
the image is converted to a density map.  Further smoothing is done
with an adaptive kernel in which the kernel size is inversely
proportional to the density. The center of this figure is at
4:30:32.0, -61:34:03.0 (J2000). The inner region contains two
components: a main cluster to the southwest and a secondary subcluster
cluster of about 30 galaxies to the northeast. This barlike structure
in the galaxy distribution is also seen as an elongation in the
central region of the X-ray contour maps (see the right panel of
Figure~\ref{fig:fig2} for the PSPC and the left panel of Figure~\ref{fig:fig3} for the GIS).

This number of galaxies in the secondary cluster is typical of a loose
group. The velocity distribution of the components in the central
region are almost identical.  Biweight estimates of the heliocentric
velocity are 17804 km s$^{-1}$ for the main cluster core and 17679 km
s$^{-1}$ for the secondary subcluster.  This means that using velocity
information to separate them is nearly impossible.  The peak at 18400
km s$^{-1}$ not significant with greater than 90\% confidence.  Thus,
the inner sample does not show any signs of being non-Gaussian,
suggesting that the merger is in the plane of the sky.

\subsection{X-ray}

Comparison of the 90\% confidence range on the free parameters derived
from modeling single data sets for the (PSPC:GIS) shows that while the
column density (0.023 - 0.026:0.023 - 0.039 cm$^{-2}$) and abundance
(0.0 - 0.50:0.14 - 0.25 Solar) ranges are in good agreement, the
emission weighted temperatures (4.7 - 6.9:7.7 - 8.8 keV) are not (see
Table~\ref{tab:tab1}).  The PSPC is more sensitive to the cooler gas component of
the ICM and so the emission weighted temperature measured with the
PSPC is lower than that measured by the GIS.  This is indicative of
non-isothermality in the gaseous atmosphere. A two temperature model
is marginally preferred over a single temperature model when both data
sets are fit simultaneously.  The two temperature components (0.31 -
6.5:7.82 - 12.94 keV) differ with 90\% confidence.

\begin{deluxetable}{ccc}
\footnotesize
\tablewidth{0pt}
\tablecaption{Region Temperatures}
\tablehead{
\colhead{Region Number} & \colhead{Best Fit Temperature (keV)} & \colhead{
90\% Temperature Range (keV)}}
\startdata
1 & 6.6 	& 	5.8 - 7.5 \nl
2 & 6.8 	& 	6.0 - 7.6 \nl
3 & 8.0 	& 	7.4 - 8.5 \nl
4 & 7.8 	& 	7.3 - 8.4 \nl
5 & 9.0		& 	8.6 - 9.6 \nl
6 & 9.9 	& 	9.2 - 10.8 \nl
7 & 8.8 	& 	7.8 - 9.8 \nl
8 & 13.2 	&      11.0 - 16.7 \nl
\enddata
\label{tab:tab2}
\end{deluxetable}

\begin{figure}
\centerline{\includegraphics[width=6.0in]{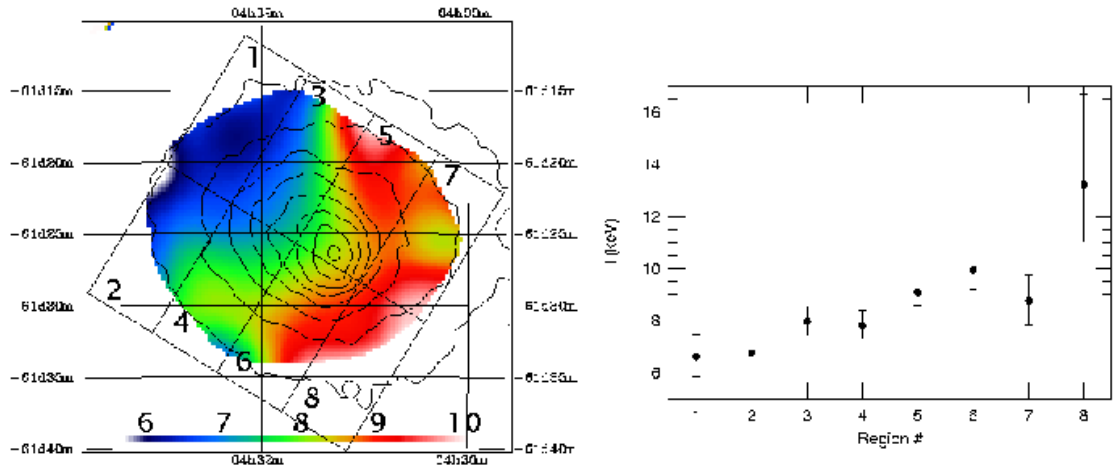}}
\caption{The left panel shows a color-coded temperature map
obtained from modeling the GIS while accounting for the PSF.  The
eight regions were chosen to cover the merger axis and give the
temperature variation across it. The surface brightness contours from
the GIS are also shown. The right panel gives the temperatures with
90\% confidence error bars for each of the regions.  The temperature
is peaked in regions 5 and 8 and generally decreases from SW to
NE. This is consistent with shock heating where the subcluster to the
NE, shown in the optical image (left panel of Figure~\ref{fig:fig2}), has
penetrated the main cluster.}\label{fig:fig3}
\end{figure}

Modeling a circular region of radius 1.04 h$_{50}^{-1}$ Mpc, while
correcting for the PSF, shows there are temperature variations across
the atmosphere (see Figure~\ref{fig:fig3}, left panel). This figure shows the color
coded temperature map obtained with the GIS with the GIS surface
brightness contours overlayed. The map represents a smoothed
temperature map in which each 15" pixel region is modeled as an
emission weighted sum of the high (9 keV) and low (3 keV) cluster
temperatures (Churazov et al. 1996). Though the errors on each
temperature are large, the general temperature structure indicates a
shock structure perpendicular to the merger axis.  To study the
temperature with greater accuracy, we have defined a set of regions
arranged to probe the temperature along the merger axis.  The
temperatures obtained for the rectangular regions are given in Table~\ref{tab:tab2}
and shown graphically in the right panel of Figure~\ref{fig:fig3}, with 90\%
confidence error bars.  We find there is an overall trend of
decreasing temperature from West to East, with the hottest gas located
west of the X-ray peak in a regions 5 and 8 where the intensity
contours are also compressed. If the overall temperature distribution
has structure on small scales, the regions chosen with which to sample
the distribution will affect the final results. For example, our
region 8 is a composite of parts of regions 2, 6, and 9 in the map
contained in Markevitch et al. (1998).  Region 2 in their analysis has
a temperature range of 6 - 12 keV and is consistent with our
measurement of region 8 (11-16.5 keV).  The temperature measured in
our specific regions are in good agreement with those found for A3266
using different regions and an independent method of correcting for
the PSF (Markevitch et al. 1998).  For a comparison of the two PSF
modeling methods see Donnelly et al. (1998).  The optical substructure
can be spatially identified with the temperature structure by noting
that the box shown on the optical contour map (left panel of Figure~\ref{fig:fig2})
is the region over which the temperature map was made (left panel of
Figure~\ref{fig:fig3}).  The distribution of galaxies, taken with the velocity
structure, suggests that either the secondary galaxy component has
just passed through the primary cluster, entering it from the SW, or
it is falling into it from the NE, prior to the initial core passage.
The indication of substantial shocked gas in our temperature maps
favors the former model, implying that the subcluster has already
penetrated the core of the main cluster along the axis running from SW
to NE. The heating appears perpendicular to the merger axis and mainly
on one side of the cluster.  In this regard it is similar to the
temperature pattern found for the Coma cluster (Honda et al. 1996).

\section{Discussion}

%Cluster mergers provide a view of how gravitationally bound structure forms. 
%Optical and X-ray analysis gives a description of the physical
%environment of a merger which shows that mergers have a significant
%affect on galaxy evolution and radio galaxy morphology.
Girardi et al. (1997) classify the A3266 optical morphology as
unimodal, having a primary cluster with a secondary cluster of
galaxies to the East.  Our optical analysis confirms this structure in
the inner core. The components have similar velocity dispersions,
which are both approximately $\sim$300 km sec$^{-1}$ higher than the
surrounding cluster. Heating in the intergalactic medium along the
W-SW rim of the cluster has resulted from collision of the cluster
atmospheres during the initial supersonic impact of the secondary
cluster falling into the primary from the SW. The morphology of the
temperature map is similar to that found by Roettiger, Burns, \& Loken
(1996) in the relatively late stage of an unequal merger (4:1 mass
ratio) 2 Gyr after core passage.

Heating of the cluster atmosphere during a merger tends to be
localized and has little effect on the emission weighted temperature.
Comparison of the emission weighted temperature derived here, 8.1 keV,
with the velocity dispersion of the outer region, 1000 km s$^{-1}$,
indicates rough energy equipartition between the gas and galaxies of
the main, pre-merger cluster.  The high inner velocity dispersion then
must result from converting kinetic energy of the infalling group, via
dynamical friction, into kinetic energy of the galaxies in the
bar-shaped inner region of the cluster.

\subsection{The Effect of Merger on the Dumbbell cD Galaxy}

Subcluster merger and accretion of galaxy groups has been observed in
several clusters (e.g., A754, A1367, A3395, A3528, Coma) consistent
with N-body simulations which show multiple mergers occurring as
clusters are assembled.  Recent studies of the Coma cluster show that
numerous mergers have taken place. Groups around each of the central
dumbbell, NGC 4874 and NGC 4889, have fallen through the main cluster
core (Donnelly et al. 1999; Gambera et al. 1997; Colless \& Dunn 1996)
and a group around NGC 4839 has penetrated the core (Burns et
al. 1995).

A3266 also has a central cD galaxy with a faint secondary nucleus in a
dumbbell morphology.  The right panel of Figure~\ref{fig:fig2} shows the smoothed,
flatfielded {\it ROSAT} PSPC image overlayed on the Digitized Sky
Survey image of the central 8.5x8.5' region of the A3266
cluster. While the cD galaxy in the dumbbell pair, located at
4:31:12.16, -61:27:15.3) appears slightly offset (30") from the X-ray
center, this may be due to a combination of smoothing and the {\it
ROSAT} pointing error, $\sim$15".  The faint nucleus and the cD have a
relative velocity of $\sim$400$\pm$39 km s$^{-1}$, somewhat larger
than the velocity dispersion of the nucleus of the cD (327$\pm$34 km
s$^{-1}$).  Carter et al. (1985) found that the cD shows a rise in
stellar velocity dispersion with radius to a maximum of 700 km
sec$^{-1}$. This rise is asymmetric and can be understood in context
of the subcluster merger. We propose that it is a sign of a tidal
interaction. The interaction appears to be with the infalling
subcluster rather than the faint nucleus. The region around the cD,
visible in the right panel of Figure~\ref{fig:fig2}, shows a tidal extension along
the merger axis in the direction of the subcluster which has passed
through the main cluster. While the velocity difference of the nuclei
is large, it is consistent with the overall rise in the velocity
dispersion with radius due to the tidal interaction.  The cD galaxy
has a heliocentric velocity of 17795 km s$^{-1}$, very close to the
value we found for the main cluster core, 17804 km s$^{-1}$. This is
consistent with the cD being at rest in the main cluster core.  Our
hypothesis is that the faint nucleus is bound to the cD, possibly
captured from the infalling subcluster.  The merger has tidally
distorted the cD galaxy in a head-on core passage by the secondary
cluster and increased the velocity dispersion in the outer halo of the
galaxy. The merger velocity we have calculated from the temperature
map (section 5.2), 1400 km s$^{-1}$, would correspond to the relative
velocities of the galaxies in the during the merger if they are both
at rest in the potential well of their respective subclusters. While
this relative velocity is much higher than the observed velocity
difference of the dumbbell, dynamical friction would increase the
internal energy of the much more massive cD at the expense of the
kinetic energy of the other smaller galaxy. As noted above, the
heating seen in the velocity dispersion of the cD is significant,
approximately a factor of two. This hypothesis would account for the
observed tidal distortion of the cD halo along the merger axis, the
previously reported rise in velocity dispersion in the halo, and the
large relative velocity of the cD and faint nucleus.

Both A3266 and Coma were included as high density environments in a
study of the effect of the dense cluster environment on star formation
history (Rose et al. 1994).  This study found that the mean spectral
type of evolved stars in early-type galaxies is considerably later and
more metal enriched in low-density regions when compared to high
density regions. This implies that star formation is essentially
truncated in the dense cluster environment at an earlier epoch. This
is consistent with studies which show that the galaxies currently
showing star formation in the Coma cluster are late infalling disk
galaxies.  Studies of the stellar content of dumbbell galaxies
(Quintana et al. 1990) are in general agreement with these results
showing that the typical stellar population is red with strong
metallic absorption lines. However, some of the dumbbells also had a
blue stellar population in their inner 3 kpc.  Since all of the pairs
are presumably subject to mutual interaction, it is unlikely that the
star formation is triggered by galaxy interaction.  Alternatively,
those systems with a star forming population may be experiencing a
larger scale environmental effect such as a merger.  Mergers may
affect the evolution of member galaxies through ram-pressure stripping
and subcluster tidal effects: both of which affect the star formation
history of galaxies.  Confirmation of this interpretation of the blue
stellar population in dumbbells may come through future optical
studies of systems such as the A3266 dumbbell since it is in the midst
of a merger and appears to show a tidal distortion.

\subsection{The Effect of Merger on Radio Morphology}

Robertson \& Roach (1990) report a 3.28 Jy source at 408 MHz
associated with the A3266 cluster. The Molonglo Reference Catalog type
is classified as extended with a WAT morphology.  The radio source
position is 7.8' from the X-ray center to the SW.  For comparison, the
subcluster and main cluster are about 15' apart. Jones \& McAdam
(1992) report three sources at the position of the A3266 galaxy
cluster in a survey at 843 MHz. The main source at 2.79 Jy and two
weaker sources nearby. A3266 also has a narrow angle tail radio galaxy
(NAT) $\sim$7 arcmin to the NE of the WAT. The location of the sources
is to the SW of the X-ray center near the edge of the hottest region.
As visible in Figure~\ref{fig:fig1} in Jones \& McAdam, the WAT opens up toward the
shock front and the NAT tail points toward the shock front. This
implies that the shock front has passed by these galaxies and bent
their jets in the direction of the post-shock gas motion.

Burns et al. (1994) suggest WAT morphologies may be identified with
cluster mergers. This is supported by Gomez et al. (1997) who found
that 90\% of WAT clusters have evidence of substructure.  Bilton et
al. (1998) found that ROSAT images of 15 Abell clusters with
narrow-angle tail radio sources have evidence of substructure. The
velocities of the galaxies themselves are not sufficient to bend the
tail into a U-shaped morphology.

Since this merger is in the plane of the sky, it is impossible to use
the relative projected separation and velocity dispersion of the
subclusters to obtain bound dynamical solutions for the merger as has
been done for a number of galaxy clusters.  Alternatively, we use
X-ray determined pre- and post-shock gas temperatures derived from our
temperature map to estimate the relative gas velocity. This
calculation has been done for several clusters (Markevitch, Sarazin,
\& Vikhlinin 1999). We use the formulations of the Rankine-Hugoniot
jump conditions contained in their paper which gives the relative pre-
and post-shock gas velocity in terms of the pre- and post-shock gas
temperature. This analysis does not assume a strong shock. Using
Figure~\ref{fig:fig3}, the pre- and post-shock gas temperatures are estimated at
6.5 and 13 keV, respectively.  Using these parameters in the shock
equations gives a relative gas velocity of 1400 km sec$^{-1}$. This
simplified physical picture is consistent with simulations that show
gas velocities are commonly in the range of 1000 - 3000 km sec$^{-1}$
(Roettiger, Burns, \& Loken 1996).  The WAT has been tentatively
identified with an 18.5 magnitude galaxy at 4:29:35.21, -61:38:10.4
(1950) by Jones \& McAdam (1992). This galaxy has a mean velocity of
17042 km s$^{-1}$ (QRS) relative to the main cluster component
velocity, 17804 km s$^{-1}$, reported in Section (4.1). Gomez et
al. (1997) used the observed radio parameters of a WAT in NAT in Abell
clusters they studied to calculate a required jet bending velocity of
$>$1000 km s$^{-1}$. The velocity of the post-shock gas we calculated,
1400 km s$^{-1}$, is nearly a factor of two higher than the
line-of-sight galaxy velocity, 766 km s$^{-1}$, which would provide 4
times the ram-pressure on the galaxy's jets. Thus, the relatively low
galaxy velocity and the higher ram-pressure provided by the merger
induced gas motion imply that the WAT morphology is due to the merger
rather than the motion of the galaxy in the ICM. This argument is
strengthened by the fact that the NAT and WAT have the same alignment,
relative to the shock front.

\section{Conclusions}

We have presented an X-ray temperature map of the A3266 galaxy cluster
made from observations obtained with {\it ASCA}. The map shows an
asymmetric heating pattern with the hottest region at
13.2$^{+3.4}_{-2.2}$ keV to the West of the X-ray brightness peak. The
galaxy kinematical data has also been partitioned into subclusters and
has two components, a main cluster and a loose group to the NE. Both
components have a similar velocity so the merger is assumed to be in
the plane of the sky. We have found ample evidence of a merger in
A3266: shocked gas, increased velocity dispersion in the central
region, asymmetric optical and X-ray morphology, and disturbed radio
galaxy structure.  The temperature constraints were used to estimate
the relative pre- and post-shock gas velocity of 1400 km s$^{-1}$.
The ram-pressure together with our reconstructucted merger geometry
can account for a WAT and NAT to the SW of the X-ray peak. The high
relative velocity between the cD its secondary faint nucleus, which
are in a dumbbell morphology, as well as the high, asymmetric velocity
dispersion of the cD is attributed to tidal heating from the recent
passage of the subcluster through the core of the main cluster.

\acknowledgements

The authors thank Eugene Churazov and Marat Gilfanov for use of their
software and Maxim Markevitch for useful comments on this manuscript.
MJH acknowledges support from the National Science Foundation grant
AST-9624716.  RHD acknowledges support from the Smithsonian
Institution and NASA grant NAS8-39073.

\clearpage

%\plotone{fin_fig1.eps}
%\clearpage
%\plotone{newcombo2.eps}
%\clearpage
%\plotone{newcombo3.eps}

\end{document}